\documentclass[prl,aps,twocolumn,floats,superscriptaddress,showpacs]{revtex4}
\usepackage{graphicx}

\newcommand{\be}{\begin{equation}}
\newcommand{\ee}{\end{equation}}
\newcommand{\bea}{\begin{eqnarray}}
\newcommand{\eea}{\end{eqnarray}}

\newcommand{\ra}{\rangle}

\newcommand{\lp}{\left(}
\newcommand{\rp}{\right)}
\newcommand{\lf}{\left\{}
\newcommand{\rf}{\right\}}

\renewcommand{\H}{{\cal H}}

\newcommand{\E}{{\cal E}}

\renewcommand{\P}{{\cal P}}

\newcommand{\tr}{{\rm tr }\,} 

\begin{document}
\title{Energy Anomaly and Polarizability of Carbon Nanotubes}
\author{D. S. Novikov}
\affiliation{Department of Electrical Engineering and Department of Physics,
Princeton University, Princeton, NJ 08544}
\author{L. S. Levitov}
\affiliation{Department of Physics, Massachusetts Institute of Technology, 
Cambridge, MA 02139}

\date{\today}

\begin{abstract}
The energy of electron Fermi sea
perturbed by external potential, represented as energy anomaly
which accounts for the contribution of the deep-lying states, is analyzed 
for massive $d = 1+1$ Dirac fermions 
on a circle.
The anomaly is a universal function of the 
applied field,  
and is related to known field-theoretic anomalies.
We express transverse polarizability 
of Carbon nanotubes via the anomaly, in a way which exhibits 
the universality and scale-invariance of the response
dominated by $\pi$ electrons and qualitatively different 
from that of dielectric and conducting shells.
Electron band transformation 
in a strong-field effect regime is predicted.
\end{abstract}


\maketitle

Fermion anomalies are universal contributions 
to low-energy properties of field theory
originating from the bottom of the filled Fermi-Dirac sea.
The primary examples in high-energy physics are   
the spontaneously generated photon mass in $d=1+1$ 
\cite{Schwinger}, and 
the Adler-Bell-Jackiw chiral anomaly \cite{ABJ} 
relating the decay of pion into two photons to 
the number of quark colors \cite{Peskin}. 
Several instances of anomalies are known in 
solid state physics,
with the chiral anomaly manifest in transport \cite{anom-cm-chiral}
and in fermion number fractionalization \cite{frac}, 
and the parity anomaly in $d=2+1$ linked 
to the quantum Hall effect \cite{anom-cm-parity}.

Here we describe a new manifestation of fermion anomaly, appearing
in the total energy of Fermi sea in the presence of an external 
field. 
While the net energy of Fermi-Dirac vacuum depends on 
ultraviolet cutoff, i.e. on the behavior
at the bottom of the band, we find 
the external field-dependent part of the energy
to be cutoff-insensitive.
The latter energy is naturally
divided into two parts, 
one given by a sum of the energy shifts of filled
states near the Fermi level, and another, equally important,
desribing cumulative effect of the states deep below the Fermi 
level. The latter, anomalous part, 
traced to Schwinger anomaly \cite{Schwinger},
has a universal form and can be expressed through 
the properties near the Fermi level.

The Fermi sea energy field-dependence is the physical quantity central for 
many physical properties of materials.
As an application, below we consider the response of 
the carbon nanotube (NT) $\pi$ electron band to
perpendicular electric field. 
The aforementioned separation into the normal and anomalous parts
enables one to handle energy in a fully general and, at the same time,
case-specific way, taking full account of the level quantization, 
curved geometry, spatial inhomogeniety, etc. 
The NT $\pi$-band
is described by a tight-binding model on a honeycomb lattice 
\cite{Dresselhaus}. 
Near the Fermi level, at the $\pi$-band center, the NT electrons
are described by $2+1$ Dirac fermions moving on a cylindrical 
surface \cite{DiVincenzo-Mele}. Dirac model provides a simple 
analytic description
of NT curvature and chirality  \cite{Dresselhaus,curvature}, and of the effects
of external fields \cite{B-paral} in both semiconducting and metallic NTs.

It might seem that understanding properties such as NT 
polarizability should require a detailed knowledge of the 
$\pi$ electrons behavior on the lattice constant scale 
\cite{Benedict95}.
Here we find that, to the contrary, the energy calculated from Dirac model,
with the anomaly properly accounted for, not only is numerically 
accurate, but also provides new insight into NT properties. 
We identify the dominant role of $\pi$ electrons
response compared to that of other orbitals,
and explain the origin of the scale-invariant depolarization,
independent of the NT diameter \cite{Benedict95,NovikovLevitov,Rotkin},
qualitatively different from that of metallic and dielectric shells.
Our approach, with
electron interactions included in an RPA fashion 
\cite{Gonzalez99KaneMele04},
is not limited to linear response: we 
apply it to study NT electron band transformation in the field-effect regime.

The origin of the energy anomaly is exhibited most clearly by the example of 
chiral 1d fermions on a circle $[0,2\pi]$, described by the Hamiltonian
$\H = -i\partial_\theta + U$, $U = 2a\cos\theta$.
In the Fourier representation $|n\ra=e^{in\theta}$, $\H$ is given 
by an infinite three-diagonal matrix:
$\H_{n\,n}=n$, $\H_{n\,n\pm 1}=a$. 
Eliminating the potential $U$
by a gauge transformation
$\psi(\theta)=e^{-2ia\sin\theta} \widetilde\psi(\theta)$, 
$\widetilde{\cal H} = -i\partial_\theta$, we see that
the eigenvalues of $\H$ are integers independent of $U$.

The energy anomaly arises when 
the interaction $U$ is {\it truncated} at a certain energy scale.
To that end, let us consider a more general three-diagonal matrix
\be \label{eq:3-matr}
\H_{n\,n}=n,\quad \H_{n\,n+1}=\H_{n+1\,n}=a_n,\quad n\in Z,
\ee
with the sequence $a_n$ 
having different limits at $n\to\pm\infty$: $a_{n\to-\infty}=0$,
$a_{n\to+\infty}=a$.
Although now the energy levels depend on $a_n$, 
the above 
argument for spectrum robustness at constant $a_n$ indicates that
this dependence is exponentially 
small at large $|n|$. The level shifts 
are significant for only a relatively 
small cluster of levels around $n\simeq n_\ast$
where the switching of $a_n$ from $0$ to $a$
occurs.

Notably, the sum of all level shifts, 
$\delta \tr \H = \sum_n \delta \epsilon_n$,
depends only on the asymptotic values $a_{n\to\pm\infty}$,
while other details of the sequence $a_n$ do not matter.
To see this, we truncate the 
matrix $\H$ at some large positive and negative $n=\pm N$,
in which case the trace of this $(2N+1)\times(2N+1)$ matrix
is finite and explicitly $a$-independent.
Since at $N\gg n_\ast$, 
the mutual influence of the levels at $n\simeq n_\ast$ and $n\simeq N$ 
is exponentially small, the effect 
on $\tr \H$ due to truncating at $n\simeq N$ is 
negative of that due to $a_n$ switching at $n\simeq n_\ast$, both being universal.
[The levels at $n\simeq -N$ are unaffected by the truncation since 
$a_{n\to-\infty}=0$ and 
$\H$ is diagonal.] 
Thus the sum 
of the level shifts at $n\simeq n_\ast$ depends only on $a_{n=+\infty}=a$,
giving a cutoff-independent contribution to $\delta \tr \H$.

The universal value $E_{\rm anom}=\delta\tr\H$ can be evaluated using
slowly varying $a_n$,
$|da_n/dn|\ll |a_n|$. In this case,
since the levels are unperturbed by constant $a_n$, 
the level shifts will be small,
which warrants using pertubation theory.
Gradient expansion of $a_n$ in the vicinity of $n=n_\ast$
yields ${\cal H} = -i\partial_\theta + 2\bar a\cos\theta 
+ b \lf e^{i\theta} (-i\partial_\theta) + {\rm h.c.} \rf$,
with 
$\bar a=a_{n_\ast}-bn_\ast$,
$b=da_n/dn|_{n=n_\ast}$.
The gauge transformation 
$\psi(\theta)=e^{-2i\bar a\sin\theta} \widetilde\psi(\theta)$
transforms $\H$ into
\be
\widetilde{\cal H} = -i\partial_\theta 
+ b\lf e^{i\theta}(-i\partial_\theta - 2\bar a\cos\theta)  + {\rm h.c.} \rf .
\ee
The energies $\epsilon_n$ with $n$ near $n_\ast$ obtained in the lowest order 
of perturbation theory are
$\epsilon_n = \langle n| \widetilde{\cal H} |n\rangle = n-2b\bar a$.
The sum of these level shifts, given by a full derivative,
depends only on the asymptotic of $a_n$:
\be\label{eq:Eanom=-a^2}
E_{\rm anom}
= -\sum_n 2b\bar a
\simeq -\int_{-\infty}^{\infty} \!dn\, 2a_n \frac{da_n}{dn} = -a^2 \,.
\ee
The relation of this result with fermionic energy
emerges when one considers quenching of the external field 
$a_n=a$
coupling to the states at the Fermi sea bottom,
modeled by $a_{n\lesssim n_\ast}=0$.
We find that the anomalous contribution $E_{\rm anom}$
is universal, i.e. it depends only on the properties 
near Fermi level \cite{footnote1}.
The anomaly contributes additively to the energy along with
the contributions due to fermion mass and confinement (see below).

Electrons in a nanotube, a cylinder of radius $R$, are described by $2+1$ 
massless Dirac model \cite{DiVincenzo-Mele}. 
The states in a transverse field, labelled 
by momentum $k$ along the tube, can be viewed as 
massive $1+1$ Dirac fermions on NT circumference, 
the circle $0<y<2\pi R$: 
\be
\label{H-dirac}
\H_D=-i \hbar v \hat \alpha \partial_y + \hbar v k \hat\beta  + U(y) \,,
\ee
with $U(y)$ the transverse field potential 
and $\hbar v k$ the Dirac ``mass''.
(Here $\hat\alpha=-\sigma_x$ and $\hat\beta=\sigma_y$ \cite{Dresselhaus}.)
We assume generic quasiperiodic boundary conditions
$\psi(y+2\pi R) = e^{2\pi i \delta}\psi(y)$. 
At $U\equiv 0$, the energy levels 
\[
\epsilon_n^{\pm} = \pm\sqrt{\Delta_R^2 (n+\delta)^2 + (\hbar v k)^2} \,,
\quad \Delta_R \equiv \hbar v/R \,,
\]
describe NT minibands. 
The phase $\delta$ determines NT properties: 
$\delta=\pm\frac13$ for semiconducting NT, $\delta=0$ for metallic NT,
and $|\delta|\ll 1$ for the tubes 
with small gap induced by curvature \cite{curvature} 
or by parallel magnetic field \cite{B-paral}.

For the Fermi sea energy at finite $U$, the naive answer would be the sum
over occupied states
$W = \sum \delta \epsilon^{-}_n$, 
$\delta \epsilon_n \equiv \epsilon_n|_U - \epsilon_n|_{U=0}$.
The level shifts $\delta \epsilon_n$ become very small 
away from the band center,
at $|n|\gg kR$: In this limit, 
with the contribution of finite mass being negligible,
the Dirac problem (\ref{H-dirac}) decouples into {\it two} chiral fermion modes,
each having $U$-independent spectrum.
However, despite the absence of level shifts, the states with large
$n$ contribute to the $U$-dependent energy via anomaly
due to the bandwidth cutoff,
$(\hbar v/R)n_\ast\sim -10\,{\rm eV}$ for Carbon.
As described above, 
the anomaly depends only on the properties near the band center:
\be\label{E-anom}
E_{\rm anom} = - \frac{N_f}{2\pi \hbar v} \int_0^{2\pi R}\!\!\! U^2(y)dy \,,
\ee
where $N_f=4$ is the number of 
electron flavors associated with spin and the points $K$, $K'$ \cite{Dresselhaus}.
Note that 
the energy (\ref{E-anom}) is {\it additive} 
for multiple fermion flavors, contrary to fermion-doubling cancellation 
typical of the chiral anomaly effects \cite{fermion-doubling}.

One can interpret Eq.(\ref{E-anom}), somewhat loosely, as a counterterm
which eliminates the nonphysical contribution
of infinite Fermi sea in the model (\ref{H-dirac}), i.e.
the states outside the Carbon band.
A photon mass $m_{\gamma}^2 = e^2/\pi$ appears
in $d=1+1$ QED \cite{Schwinger}
after integration over massless fermions.
Eq.(\ref{E-anom}) 
exhibits a similar effect in the 
Dirac system in an external electromagnetic field 
$e A_\mu = (U(y), 0)$, with the mass term
$\int\! d^2x \, \frac12 m_\gamma^2 A_\mu^2 \equiv  - \int \! E_{\rm anom} dt$
in the action. 
[We point out a distinction of
our approach and Coleman's analysis \cite{Coleman76} 
of massive Schwinger model in external field,
which yields an effective action 
containing the field intensity rather than potential.]

The energy anomaly is related to the chiral anomaly in $d=1+1$,
since in this case the chirality $\gamma^5 = \hat \alpha$ 
enters the Hamiltonian (\ref{H-dirac}).
Formally this
can be seen via the change in the functional measure 
\cite{Fujikawa}.
First note that a chiral gauge transformation [$\hbar=v=1$]
\be \label{gauge}
\psi(y) = e^{-i\gamma^5 \phi(y)}\tilde\psi(y) \,, \quad 
\phi = \int^y \! dy'\, U(y') \,,
\ee
preserves the boundary conditions on $\psi$, turning (\ref{H-dirac}) 
into
\be \label{H-dirac-gauged}
\tilde \H_{D} = -i \hat\alpha \partial_y 
+ k e^{2i\hat\alpha\phi(y)}\hat\beta \,. 
\ee
Consider now the variation of the background field $U(y)$ by $\delta U(y)$,
related to the infinitesimal transformation (\ref{gauge}), 
$\psi' = e^{i\gamma^5 \lambda(y)} \psi$,
$\lambda(y) \equiv \int^y \!dy' \, \delta U(y')$.
The corresponding Jacobian $J = \exp\{2i \int\! dtdy\, {\cal A}\lambda(y)\}$  
\cite{footnote2}  
changes the action by 
$-i\ln J = - \delta E_{\rm anom}t$,
yielding $\delta E_{\rm anom}=\frac1\pi \int \! dy\, \lambda \partial_y U$,
with 
${\cal A} =  
\tr \gamma^5
= {1\over 2\pi}\epsilon^{\mu\nu}F_{\mu\nu} = - \partial_y U(y)/2\pi$ 
the 2d anomaly \cite{Peskin}.
Integration by parts gives the variation
$\delta E_{\rm anom} = -\frac1\pi \int\!dy\, U \delta U$,
which is equivalent to the result (\ref{E-anom})
\cite{footnote3}. 
This derivation can be extended to include electron interactions,
proving robustness of the anomaly.

Turning to the NT response to
transverse electric field,
\be \label{U}
U(y) = -e{\cal E}R \cos (y/R) \,, 
\ee
we relate polarizability to the sum 
of Stark shifts
\be\label{E0}
E_0(k)=\sum\limits_{n=-\infty}^{+\infty} \delta \epsilon_{n}(k)
\ee
$\delta \epsilon_n(k) = \epsilon_n(k)|_\E - \epsilon_n(k)|_{\E=0}$,
taken over all occupied states,
with $k$ the electron momentum along the tube.
We obtain the shifts $\delta \epsilon_n(k)$ 
from the Hamiltonian (\ref{H-dirac}).
In this calculation, 
the anomaly (\ref{E-anom}) must be added
to account for the finite band cutoff,
formally absent in (\ref{H-dirac}).

The main effect of electron interaction is depolarization, i.e.
screening of the field inside NT. 
To obtain the RPA screening function \cite{Benedict95,Gonzalez99KaneMele04} 
of NT cylinder,
we first show how the problem is reduced to the calculation
of electron energy in the presence of an external field. 
The Gauss's law relates the fields inside and outside the tube with the 
induced surface charge density $\sigma$,
\be\label{eq:Gauss}
{\cal E}_{\rm ext}={\cal E}+{\textstyle \frac12} \cdot 4\pi \cdot N_f\sigma 
,\quad
\sigma=\P/(\pi R^2),
\ee
where $\P$ is the dipole moment per flavor
and per unit NT length, 
and the factor $1/2$ accounts for depolarization in cylindrical geometry.
In Eq.(\ref{eq:Gauss}) we projected the actual charge density on 
the $\cos\varphi$ harmonic, $\varphi\equiv y/R$,  
as $\sigma(\varphi) \to N_f\sigma \cos\varphi$,
ignoring higher order harmonics. 
The problem is then reduced to evaluating the dipole moment, 
given by $\P=- dW({\cal E})/d{\cal E}$,
where $W({\cal E})$ is one fermion flavor
energy as a function
of the inner field, 
\be \label{eq:Wintegral}
W = N_f^{-1}\int_{-\infty}^\infty \lf E_0(k)+E_{\rm anom}\rf {\frac{dk}{2\pi}} \,.
\ee
Combining this with Eq.(\ref{eq:Gauss}),
and using dimensionless  $u=e{\cal E}R/\Delta_R$, we obtain 
$u_{\rm ext}=u + 2N_f \textstyle{\frac{e^2}{\hbar v}} \P(u)$.
Once the dipole moment $\P(u)$ is known,
this relation determines the inner field $u$ in terms
of the outer field $u_{\rm ext}$.

\begin{figure}
\centerline{
\begin{minipage}[t]{3.5in}
\vspace{0pt}
\centering
\includegraphics[width=3.5in]{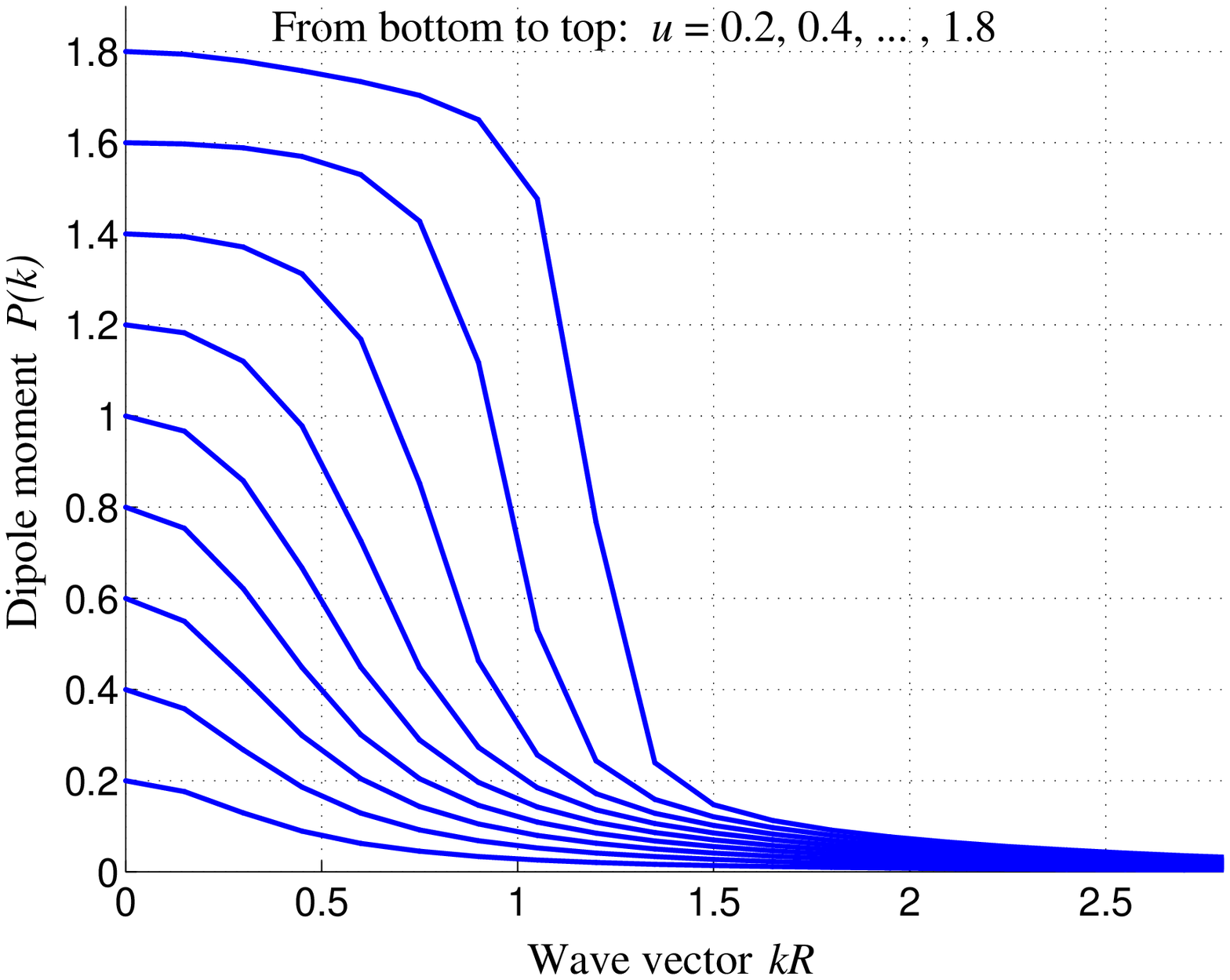}
\end{minipage}
\hspace{-1.8in}
\begin{minipage}[t]{1.7in}
\vspace{0.3in}
\centering 
\includegraphics[width=1.7in,height=1.5in]{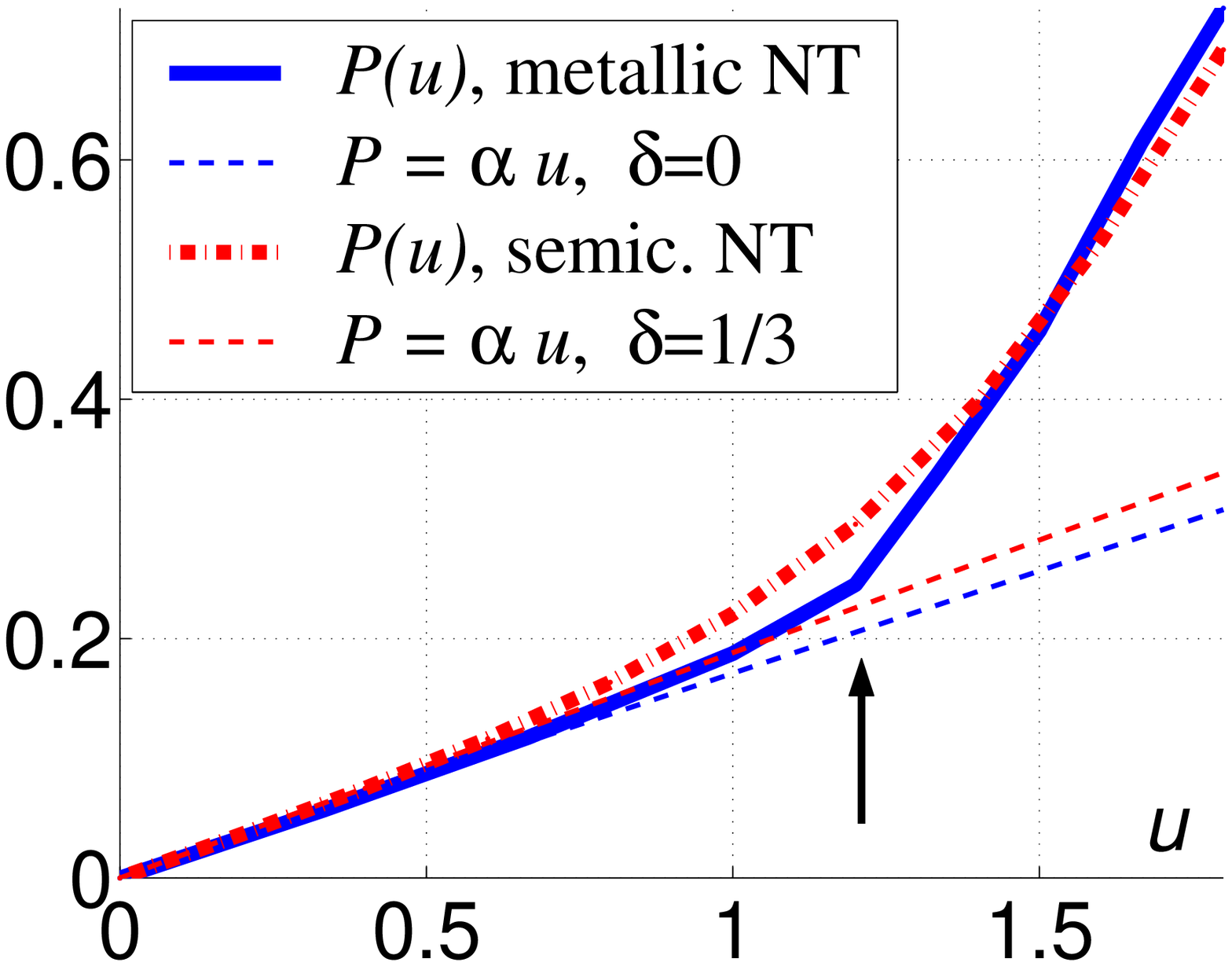}
\end{minipage}
}
 \caption[]{
Partial dipole moment $P(k)=-d(E_0(k)+E_{\rm anom})/du$ 
as a function of $k$, obtained from (\ref{eq:Wintegral})
for semiconducting NT, where $u=e{\cal E}R^2/\hbar v$
is dimensionless field.  
Note the cancellation between $E_0(k)$ and the anomaly (\ref{E-anom})
at $kR\gg1$, enforcing convergence of $\P=\int P(k)dk/2\pi$. 
Note also that $P(k\to 0)$ is dominated by the anomaly, since
$E_0=0$ at $k=0$ due to the chiral gauge invariance (\ref{gauge}).
{\it Inset:}
Dipole moment $\P$ per one fermion flavor versus $u$
for metallic and semiconducting NT. Straight lines
represent weak field linearization (\ref{Wlinearized}).
Arrow marks the field $u\simeq 1.2$ for which
velocity changes sign in metallic NT (see text and Fig.\ref{fig:MNT,SNT}).
  }
\label{fig:P(k)}
\vspace{-5mm}
\end{figure}

Taking the Stark shifts $\delta \epsilon_n(k)$ to the lowest order in $\E$
and evaluating 
the integral over $k$
(see Fig.\ref{fig:P(k)}), we obtain
\be \label{Wlinearized}
W=-\frac{\alpha}2u^2,\quad 
\alpha=\cases{0.196... & \ for $\delta=1/3$, \cr
0.179... & \ for $\delta=0$.} 
\ee
Notably, the linear relation $\P = \alpha u$
holds up to very large fields $u\sim 1$
(Fig.\ref{fig:P(k)} inset),
giving the depolarization 
%
\be\label{u-ratio}
{\cal E}_{\rm ext} = \lp 1+2N_f\alpha \, 
{\textstyle \frac{e^2}{\hbar v}} 
\rp {\cal E} \,.
\ee
With $e^2/\hbar v=2.7$ 
this gives ${\cal E}_{\rm ext}/{\cal E} = 5.24$ for $\delta=1/3$, and 
${\cal E}_{\rm ext}/{\cal E} = 4.87$ for $\delta=0$,
in excellent agreement with the full tight-binding calculations \cite{Benedict95}.

The screening (\ref{u-ratio}) 
is radius-independent and
is nearly the same in the metallic and semiconducting NTs. 
The latter is not surprising: screening 
is absent in a single 1d mode approximation, sicne polarizability 
is related to dipolar transitions between {\it different} subbands. 
The scale-invariance of (\ref{u-ratio}) 
obtained for a hollow NT cylinder
resembles depolarization in 
a massive dielectric cylinder \cite{Krcmar,Brothers05,Kozinsky}. 
The dipole moment of $\pi$ electrons, found to scale with $R^2$,
should be contrasted to $\P\propto R$ for hollow dielectric shell.
The universal scale-invariant result (\ref{u-ratio})
reflects the dominant role of $\pi$ electrons in transverse
response as compared to other Carbon orbitals.

To emphasize the role of anomaly in this calculation we note that
omitting this contribution would have led to a wrong sign 
of the response and also 
to a divergence. Indeed, due
to an upward shift of the filled levels $\epsilon^{-}_n$ (Fig.\ref{fig:MNT,SNT}), 
we have $\P_0=dE_0/du>0$, corresponding to unphysical 
``diamagnetic'' polarization sign.
Also, the $k$-dependence $E_0(k)$
causes an ultraviolet divergence in the integral
$\P=\int \! P(k)dk/2\pi$, since $E_0(k)$ increases with $|k|$, saturating
at $|k|R\gg1$ at an asymptotic value $\frac12 u^2$.
Both difficulties are resolved by taking into account
the negative 
$E_{\rm anom}=-\frac12 u^2$. 
The resulting integral (\ref{eq:Wintegral})
converges after $E_0(k)$ is offset by $E_{\rm anom}$ (Fig.\ref{fig:P(k)}).

\begin{figure}[t]
\includegraphics[width=3.3in]{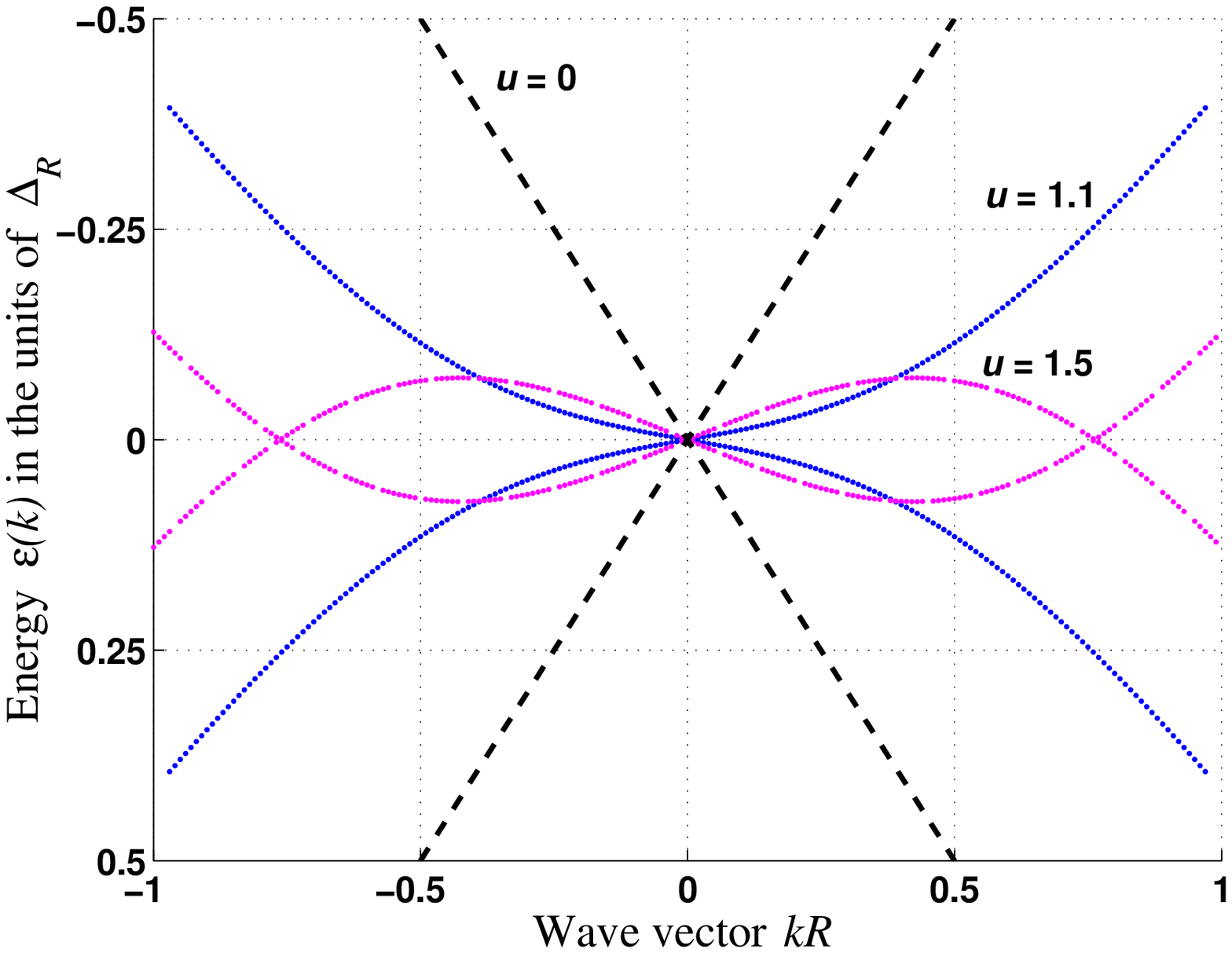}
\centerline{
\begin{minipage}[t]{3.3in}
\vspace{0pt}
\centering
\includegraphics[width=3.3in]{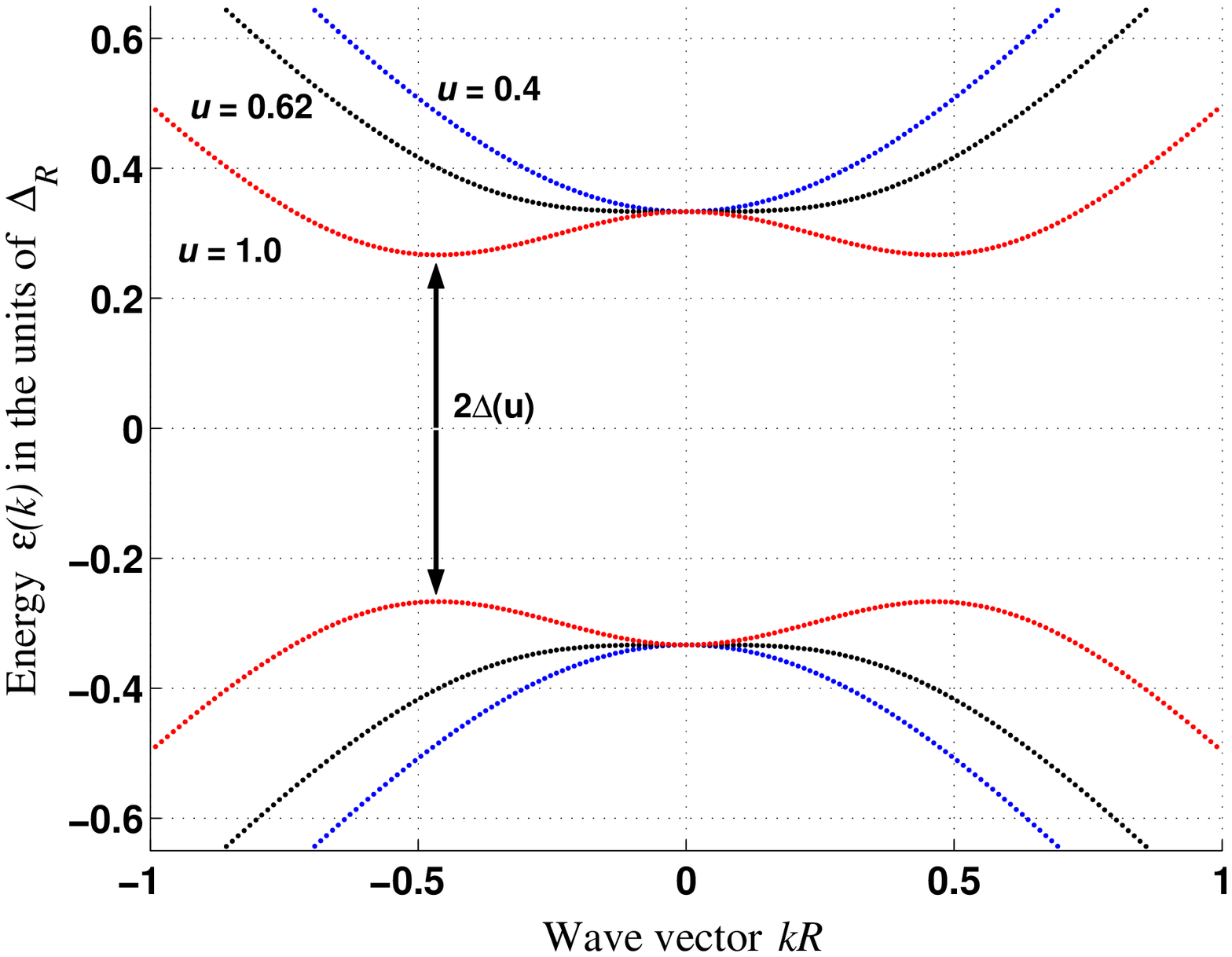}
\end{minipage}
\hspace{-1.5in}
\begin{minipage}[t]{1.4in}
\vspace{0pt}
\centering 
\includegraphics[width=1.4in]{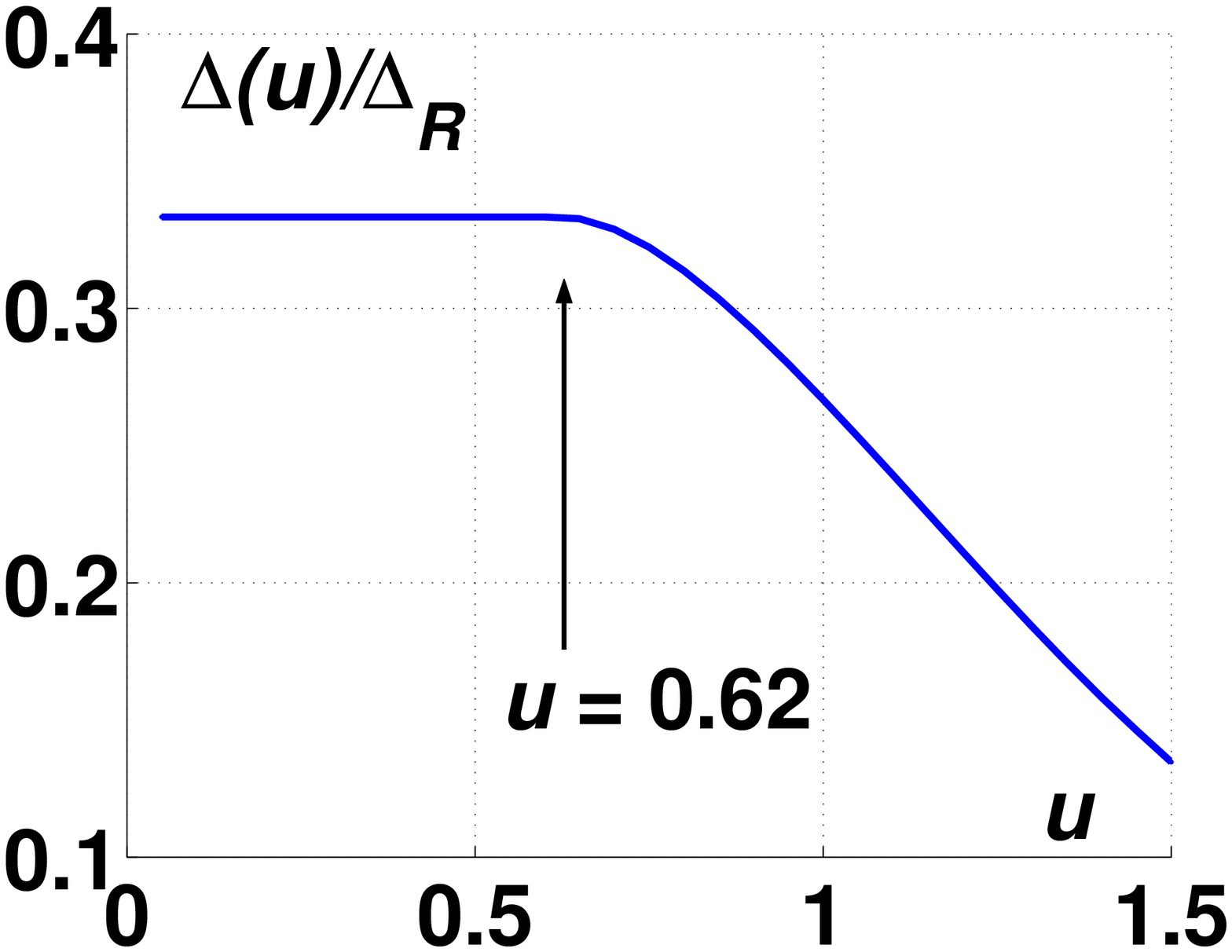}
\end{minipage}
}
\caption[]{
Electron bands transformation in the strong field-effect regime,
$u=e{\cal E}R/\Delta_R \sim 1$.
{\it Top:} 
Velocity reversal in metallic NT occurs at $u>u_c\approx 1.2$.
{\it Bottom:}
Effective mass sign change in semiconducting NT
at $u>u_c\approx 0.62$. 
The bands are shown for 
$u$ below and above the critical value. 
{\it Inset:} 
Energy gap suppression in a semiconducting NT.
  }
\label{fig:MNT,SNT}
\vspace{-5mm}
\end{figure}

To illustrate the effect of transverse field, here we
examine the NT electron spectrum. 
The changes are most dramatic in a strong field \cite{NovikovLevitov}
which mixes different NT subbands, $u \simeq 1$, 
or $e{\cal E}R \simeq \Delta_R$, 
${\cal E}\,[{\rm MV/cm}] \simeq 5.26/R^2\,[{\rm nm}^2]$.
In metallic NTs the electron velocity $v=d\epsilon/dp$
decreases and can even reverse 
sign, causing Fermi surface breakup. This could  lead 
to interesting many-body effects 
such as the Luttinger correlations increase 
due to enhanced $e^2/\hbar v$,  
or instability with respect to exciton formation for the negative-$v$ states.
Semiconducting NTs exhibit the effective mass sign change, 
accompanied by field-induced suppression of the excitation gap
(Fig.\ref{fig:MNT,SNT}) that could be manifest in activated transport. 
The above screening calculation provides an estimate for the fields
necessary to observe these effects.
The relatively small 
outer-to-inner field ratio $\simeq 5$, as well as availability 
of NTs of a few nm radius, puts the required voltage across
the tube in a feasible range of a few Volts.

In summary, we considered the energy of Fermi sea
for Dirac fermions in an external field, expressed via energy anomaly through
low-energy properties. 
The energy anomaly, applied 
to the polarizability of nanotubes,
provides insight in their response properties, 
notably the scale-invariance of depolarization,
the difference of screening in 
semiconducting and metallic NT, and relative importance of 
the $\pi$ band compared to other Carbon bands. 
Electron bands exhibit dramatic change in the 
strong field-effect regime.

This work was supported by the National Science Foundation, 
NSF-NIRT DMR-0304019 (MIT)
and MRSEC grant
DMR 02-13706 (Princeton).



\begin{thebibliography}{99}
\bibitem{footnote2}
The form of the Jacobian is due to
simultaneous change of the Grassmann variables $\psi$ and $\bar\psi$.
\bibitem{footnote3}
Note that the expression for ${\cal A}$ is consistent with 
the absence of persistent current along the circumference: 
Since only the $A_0$-component of the gauge field is nonzero and it has 
to be periodic in $2\pi R$, the Atiyah-Singer index is trivial, 
$N_R - N_L = \int \! dy dt \, {1\over 2\pi} \epsilon^{\mu\nu} F_{\mu\nu}
= \int \! dt \oint \! dy \, {\cal A} \equiv 0$.
\end{thebibliography}
\end{document}